\documentclass[conference]{IEEEtran}
\IEEEoverridecommandlockouts
\usepackage{cite}
\usepackage{amsmath,amssymb,amsfonts}
\usepackage{algorithmic}
\usepackage{graphicx}
\usepackage{textcomp}
\usepackage{xcolor}

\usepackage{subcaption}
\usepackage{url}

\def\BibTeX{{\rm B\kern-.05em{\sc i\kern-.025em b}\kern-.08em
    T\kern-.1667em\lower.7ex\hbox{E}\kern-.125emX}}
\begin{document}

\title{AI-based 3-Lead to 12-Lead ECG Reconstruction: Towards Smartphone-based Public Healthcare
}


\author{
  A. Mallick\thanks{\textsuperscript{1}Indian Institute of Technology Madras, Tamil Nadu, India.}\textsuperscript{1},
  Rahul L.R.\thanks{\textsuperscript{2}Indian Institute of Technology Hyderabad, Telangana, India.}\textsuperscript{2},
  A. Shaiju\thanks{\textsuperscript{3}National Institute of Technology Calicut, Kerala, India.}\textsuperscript{3},
  S.D. Neelapala\textsuperscript{2},
  L. Giri\textsuperscript{2}
  R. Sarkar\thanks{\textsuperscript{4}Medway NHS Foundation Trust, London, UK.}\textsuperscript{4},
  S. Jana\textsuperscript{2}
}

\maketitle

\begin{abstract}

Clinicians generally diagnose cardiovascular diseases (CVDs) using standard 12-Lead electrocardiogram (ECG). However, for smartphone-based public healthcare systems, a reduced 3-lead system may be preferred because of (i) increased portability, and (ii) reduced requirement for power, storage and bandwidth. Subsequently, clinicians require accurate 3-lead to 12-Lead ECG reconstruction, which has so far been studied only in the personalized setting. When each device is dedicated to one individual, artificial intelligence (AI) methods such as temporal long short-term memory (LSTM) and a further improved spatio-temporal LSTM-UNet combine have proven effective. In contrast, in the current smartphone-based public health setting where a common device is shared by many, developing an AI lead-reconstruction model that caters to the extensive ECG signal variability in the general population appears a far greater challenge. In this direction, we take a first step, and observe that the performance improvement achieved by a generative model, specifically, 1D Pix2Pix GAN (generative adversarial network), over LSTM-UNet is encouraging.
 
\end{abstract}



\begin{figure}

\centering
\begin{subfigure}{0.44\textwidth}
\includegraphics[width=\textwidth]{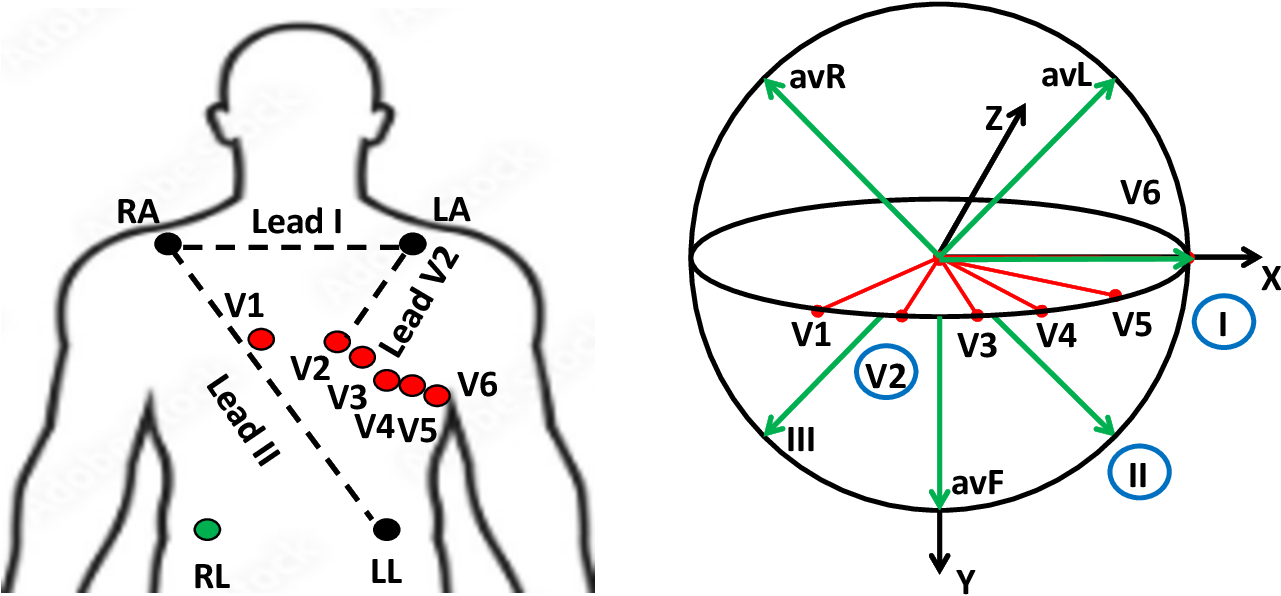}
\caption{}
\label{fig:figure2}
\end{subfigure}

\begin{subfigure}{0.5\textwidth}
\includegraphics[width=8.6cm,height=8.5cm]{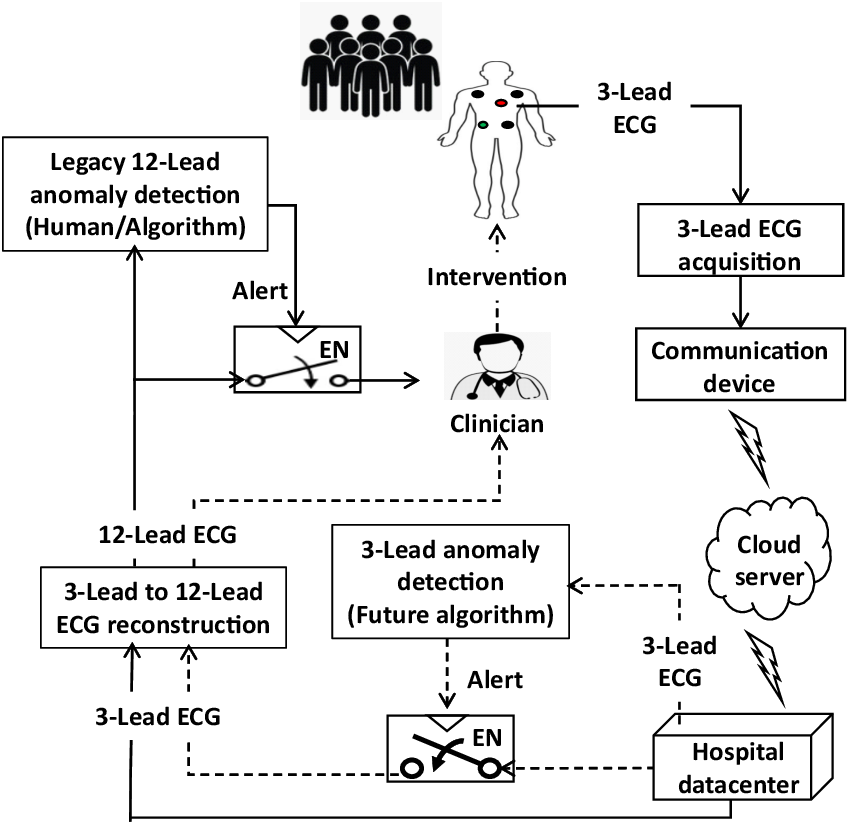}
\caption{}
\label{fig:figure1}
\end{subfigure}


\caption{(a) Standard 12-Lead and reduced 3-Lead ECG system (left); lead vectors of Standard 12-lead ECG system with the 3 leads in the reduced system circled (right). (b) Mobile public health system: Envisaged block schematic.}
\label{general_monitoring}

\end{figure}

\section{Introduction}

The electrocardiogram (ECG) remains a cornerstone of diagnosis and management of cardiovascular diseases (CVDs), and hence that of public healthcare services in view of high prevalence of CVDs. While most clinicians are trained to diagnose based on the standard 12-Lead ECG system \cite{12lead}, such a bulky system may not be ideal for mobile health services. Instead, a reduced 3-lead system (Figure \ref{general_monitoring}(a)) is preferred because of (i) increased portability, and (ii) reduced requirement for power, storage and communication bandwidth \cite{R3lead}. See Figure \ref{general_monitoring}(b) for a block schematic of a mobile public health system envisaged to collect 3-lead ECG data from remote members of the public, communicate those to medical data centers, and serve the same data in the desired 12-lead format to the clinician when a cause for medical concern (alert) arises. Of course, the 3-lead to 12-lead ECG reconstruction should be sufficiently accurate for the envisaged system to be reliable. 
Unfortunately, the existing lead-reconstruction literature rarely mentions the aforesaid public setting, while reporting extensively on algorithms in a personalized setting, where a user possesses a personal device for exclusive use and the embedded algorithms are trained on past personal data with the goal of dealing with future personal data. Examples utilize model-based as well as machine learning methods \cite{dower,affine,amit,lstm,lstm-unet}.

\begin{figure*}


\centering
\includegraphics[width=2\columnwidth]{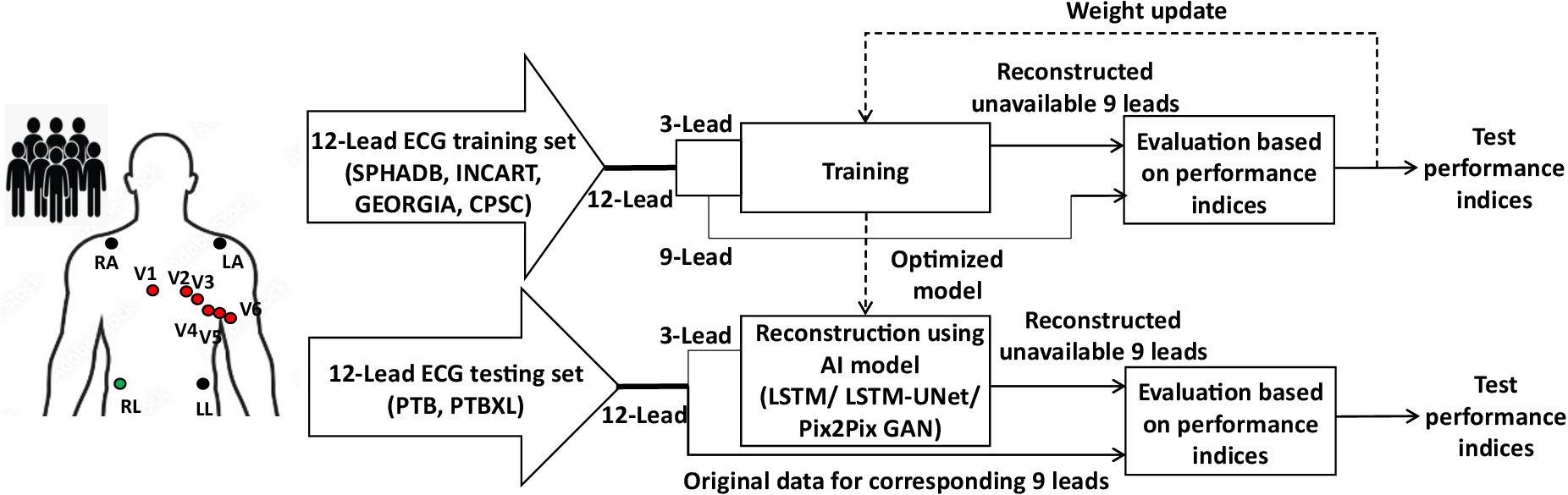}

\caption{Proposed 3-Lead to 12-Lead ECG reconstruction for public healthcare.}
\label{flowchart}

\end{figure*}

\begin{figure}[t!]

\centering
\begin{subfigure}{0.48\textwidth}
\includegraphics[width=\textwidth]{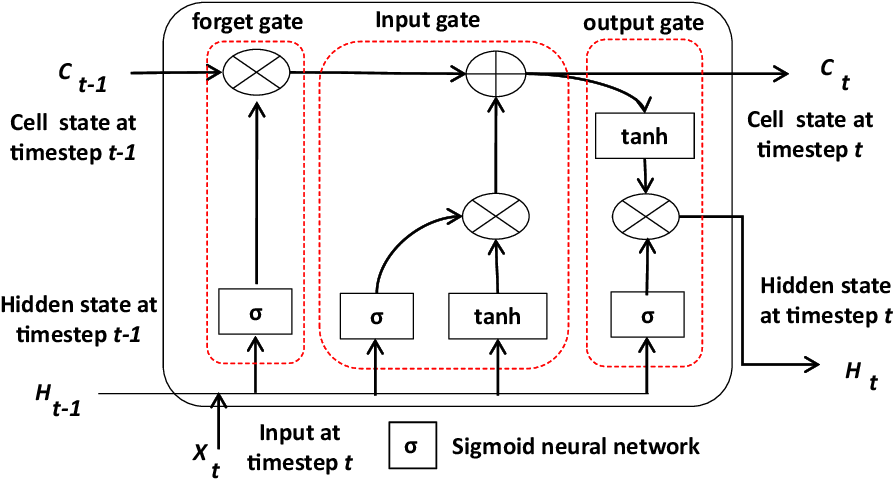}
\caption{}
\label{fig:figure2}
\end{subfigure}

\begin{subfigure}{0.48\textwidth}
\includegraphics[width=\textwidth]{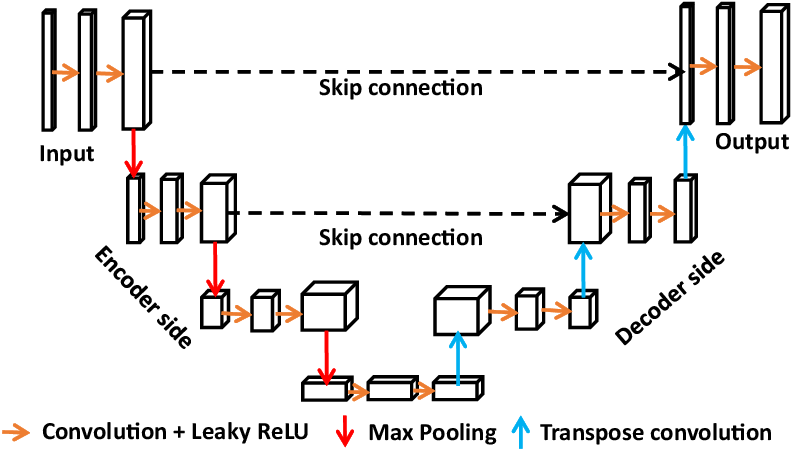}
\caption{}
\label{fig:figure1}
\end{subfigure}

\begin{subfigure}{0.31\textwidth}
\includegraphics[width=\textwidth]{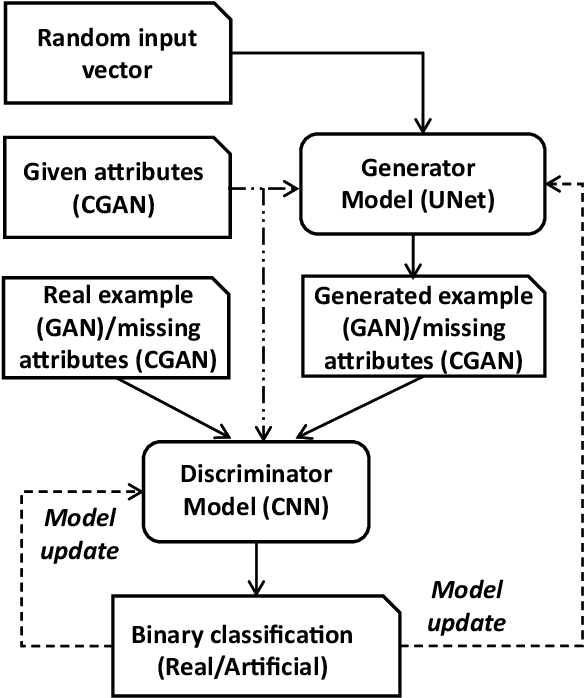}
\caption{}
\label{fig:figure1}
\end{subfigure}

\caption{Schematic diagram of (a) LSTM unit (b) UNet and (c) GAN/CGAN.}
\label{Threemodels}

\end{figure}

In contrast, the present public setting would require training based on data from a suitably representative subset of patients, and testing on data from unseen ones. As the ECG signal variability in the general population is expected to be considerably higher compared to that in individual subjects, learning a model to capture the former poses a significantly greater challenge.
In response, we take the first step by evaluating in the public setting the performance of algorithms considered the state of the art in the personalized setting, understand the performance gap and explore a possible path towards narrowing the gap and hence towards a satisfactory mobile public health system. In the personalized setting, spatial methods based on linear transforms have been suggested \cite{amit}, and tested on two publicly available 12-lead ECG databases, namely, PhysioNet’s Physikalisch-Technische Bundesanstalt (PTB) database and St. Petersburg Institute of Cardiological Technics (INCART)  arrhythmia  database \cite{physionet}. AI-algorithms exploiting temporal aspects, such as Long Short-Term Memory (LSTM), have since been found superior \cite{lstm}. A recent spatio-temporal method suitably combining LSTM and UNet has achieved further improvement \cite{lstm-unet}. 
In this paper, we consider a public setting, where against the performances of the aforesaid LSTM and LSTM-UNet algorithms, a more powerful AI tool based on Generative Adversarial Network (GAN) is proposed and evaluated for the ECG reconstruction task. Further, we report the results also on a larger and more comprehensive PTBXL database \cite{physionet}.

\section{Material and methods}

The specific task involves reconstructing nine uncircled leads (Figure \ref{general_monitoring}(a)), namely, III, aVR, aVL, aVF, V1, V3, V4, V5, and V6 based on the three circled leads I, II, and V2.

\subsection{Databases and data preparation}

As alluded, ECG lead reconstruction in the public setting requires models to be trained on a large dataset with a wide variety of cardiac representations. Accordingly, we made use of a training set that includes Shaoxing People’s Hospital’s patients arrhythmia database (SPHADB), database from China Physiological Signal Challenge in 2018 (CPSC) as well as GEORGIA and INCART databases \cite{physionet}. As depicted in the flowchart given in Figure \ref{flowchart}, the developed Pix2Pix GAN models were tested and compared on both PTB and PTBXL databases against LSTM and LSTM-UNet performances in the public setting. However, as the records (of about 5s duration) in PTBXL database are too short to train personalized algorithms, comparison of public models with the latter is feasible only for the PTB database (with 30s records). Further, all PTB-based results were presented along with subgroup analysis, using records from healthy controls (HC) and five diseased categories, namely, bundle branch block (BB), hypertrophy, cardiomyopathy and heart failure (HY), myocardial infarction (MI), valvular myocarditis and miscellaneous (VA), and those without diagnostic data (ND). Note that PTB records had been sampled at 1000 Hz, while records from the other databases at 500 Hz; hence, the former were resampled to 500 Hz for the sake of uniformity. In addition, all records were baseline corrected and normalized to the range [-1,1] \cite{sandeep}.


\subsection{Long short-term memory (LSTM)}

Traditional neural networks could be inadequate in processing temporal data because those are primarily designed for feedforward tasks, where the input/output usually does not have a sequential sense. In contrast, the current data point is influenced by previous data points in ECG time series data. In this context, the Recurrent Neural Network (RNN), proposed to capture temporal dependencies, is known to suffer from the vanishing/exploding gradient issue, hindering the capability to retain and propagate long-term dependencies. As a remedy, LSTM (Figure \ref{Threemodels}(a)) were introduced. The inherent memory cells and gating mechanisms allow LSTM models to learn and preserve relevant information, enabling them to effectively capture temporal patterns and intricate relationships between consecutive ECG samples. This ability is essential in reconstructing missing ECG leads, thus producing accurate representations of the underlying cardiac activity.

\subsection{U-shaped Network (UNet)}

U-shaped Network (UNet) represents a deep learning model,
finding extensive use in medical image analysis, which allows information fusion at different resolutions/scales (Figure \ref{Threemodels}(b)). It consists of contracting (encoder) and expanding (decoder) paths with skip connections at various levels/scales. The encoder captures the context, extracts features from the input data, and gradually reduces the spatial resolution while increasing the number of feature channels. The decoder, in contrast, fuses information inherited from the encoder with that available via skipped connections at each scale to reconstruct an output at a higher resolution. This multiscale operation allows of UNet to consider both local and global contexts within the ECG signal, and enhances its ability to accurately reconstruct missing leads.  

\subsection{Generative Adversarial Network (GAN)}

Generative Adversarial Networks (GANs) are designed to produce new statistically-indistinguishable samples from a given collection of data by training two neural networks, a generator and a discriminator, in an adversarial process (see Figure \ref{Threemodels}(c)). Taking random noise input, the generator proposes a candidate sample, while the discriminator evaluates whether that is authentic (i.e., statistically similar to the given collection). The process of training involves the generator attempting to produce increasingly realistic samples, which the discriminator attempts to detect as artificial with increasing sophistication. This adversarial process continues till the generator produces artificial data that are not reliably distinguishable by the discriminator from real ones. A related class of problems exist, where each data point consists of multiple attributes, of which a few are given and the rest needs to be recovered based on the conditional distribution of the missing ones given the available ones. Such problems are solved using a variant, conditional GAN (CGAN), where, referring to Figure \ref{Threemodels}(c), the input now consists of the given attributes along with random noise, and the real example is replaced by the available values of the missing attributes during the training phase.    


Now consider the seemingly unrelated problem of translating a black-and-white picture to a colored one. This and similar image translation problems, where one seeks missing image attributes (such as color) of each pixel from the given ones (grey scale intensity), may be posed as below: Draw samples from the conditional distribution of the desired attributes given the known ones. A variant of CGAN, called Pix2Pix GAN, has proven useful in solving image translation problems \cite{ganconditional}. Returning to the task of 3-lead to 12-lead ECG reconstruction, it poses the analogous problem of inferring 9 missing leads/attributes given the 3 given ones at every point of the one-dimensional (1D) time sequence (analogous to 2D pixel) \cite{gantimeseries}. For the current problem, noting the above analogy with image translation problems, we therefore make use of a 1D version of the 2D Pix2Pix\footnote{Although pixel stands for a 2D picture element, we shall retain this nomenclature even in 1D for the sake of notional continuity and comprehensibility.} GAN.




Specifically, we make use of a UNet-based generator and a Markovian patch/interval-based discriminator that classifies small overlapping 1D intervals (analogous to 2D patches) one at a time instead of the whole time series. Overall decision is taken via a majority principle. This approach allows the discriminator to focus on local structures rather than global features. The present
discriminator is based on a convolutional neural network (CNN) architecture designed to include high-frequency details. This approach assumes that details between intervals are independent (realistic for distant intervals).

\begin{figure*}[t!]

\centering
\begin{subfigure}{0.42\textwidth}
\includegraphics[width=\textwidth,
height=.48\textheight
]{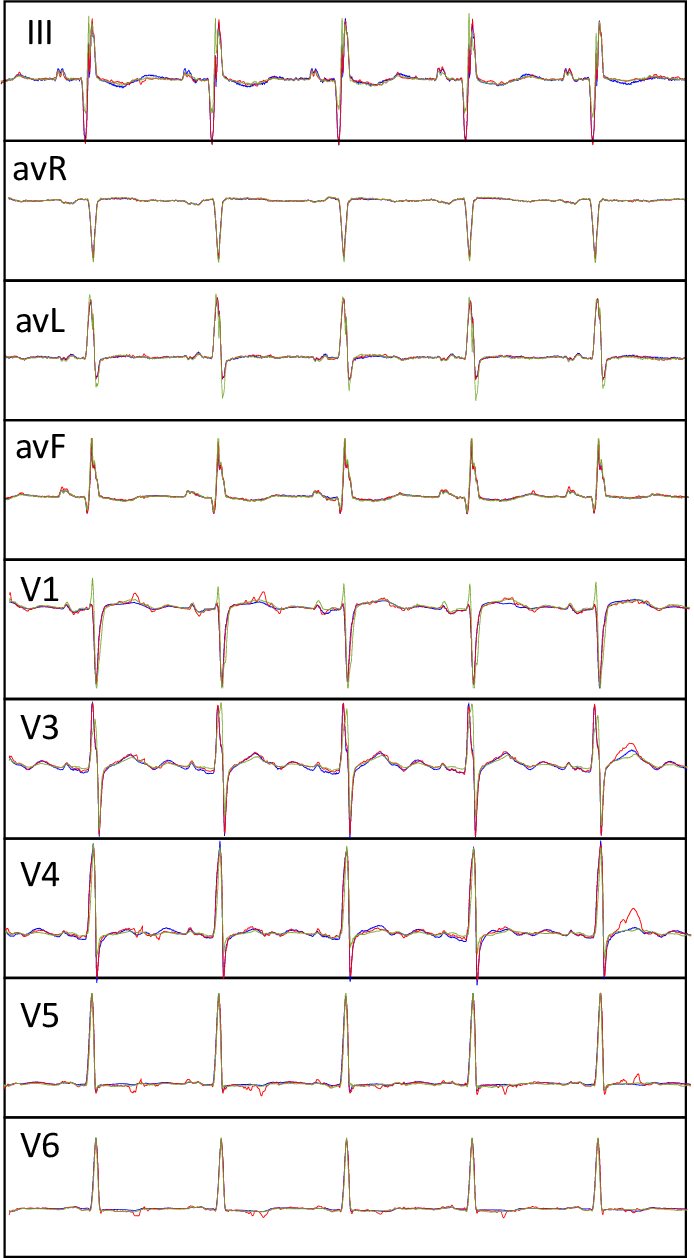}

\end{subfigure}
\quad
\begin{subfigure}{0.42\textwidth}
\includegraphics[width=\textwidth,
height=.48\textheight
]{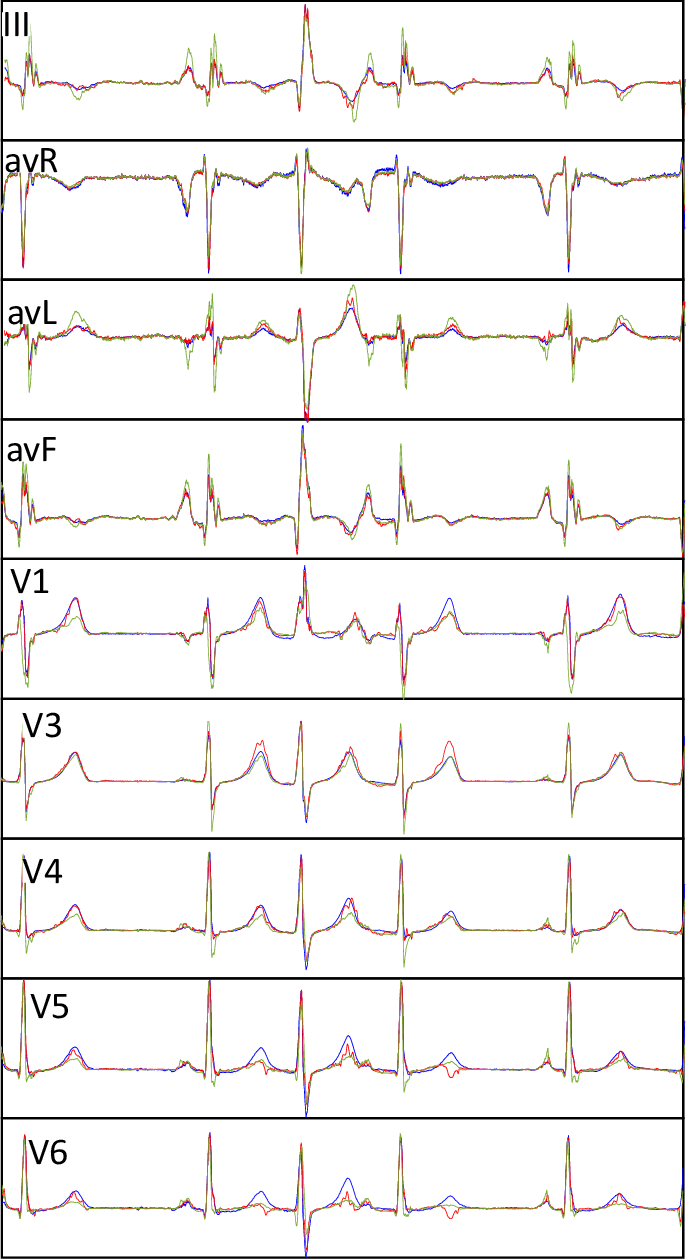}

\end{subfigure}

\caption{Each panel (with slightly altered aspect ratio) corresponds to an MI subgroup patient from the PTB (Left: Record No. patient195/s0337; Right: Record No. patient280/s0535). Blue: Original ECG; Green: Reference reconstruction using LSTM-UNet in personalized setting; Red: Proposed reconstruction using 1D Pix2Pix GAN in the public setting.}
\label{HC_UH}

\end{figure*}

\begin{table*}[t!]
\begin{center}
\begin{footnotesize}
\renewcommand{\arraystretch}{1.2}
\renewcommand{\tabcolsep}{6.5pt}

\caption{Average $R^2$ and $r_x$ between the original and reconstructed ECG leads in the public setting for patients in PTBXL and PTB. Column `a': LSTM model \cite{lstm}; Column `b': LSTM-UNet model; Column `c': Pix2Pix GAN model and Column `d': LSTM-UNet model in the personalized settings \cite{lstm-unet}, respectively.}

\begin{tabular}{|c|ccc|ccc|cccc|cccc|}
\hline
\cline{2-15}
& \multicolumn{6}{c|}{PTBXL} 
& \multicolumn{8}{c|}{PTB} \\

\cline{2-15}
& \multicolumn{3}{c|}{$R^2$}
& \multicolumn{3}{c|}{$r_x$} 
& \multicolumn{4}{c|}{$R^2$}
& \multicolumn{4}{c|}{$r_x$} 
\\
\cline{2-15}

Lead & a & b & c & a & b & c & a & b & c & d & a & b & c & d \\
\hline
III & 0.6050 & 0.6466 & \textbf{0.7781} &  0.7125 &  0.7442 & \textbf{0.9015} & 0.7151 & 0.7283 & \textbf{0.8878} & 0.9307 & 0.8067 & 0.8740 & \textbf{0.9508} & 0.9677   

  \\

avR & 0.8680 & 0.9452 & \textbf{0.9758} & 0.9317 & 0.9731 & \textbf{0.9917} & 0.8291 & 0.9479 & \textbf{0.9729} & 0.9820 & 0.9112 & 0.9739 & \textbf{0.9912} & 0.9977  

  \\

avL & 0.7252 & 0.7923 & \textbf{0.8311} & 0.8135 & 0.8387 & \textbf{0.9214} & 0.7119 & 0.8289 & \textbf{0.9183} & 0.9565 & 0.8605 & 0.9187 & \textbf{0.9660} & 0.9792   

  \\

avF & 0.7527 & 0.7976 & \textbf{0.9154} & 0.8498 & 0.9007 & \textbf{0.9685} & 0.7572 & 0.8558 & \textbf{0.9355} & 0.9645 & 0.8839 & 0.9298 & \textbf{0.9778} & 0.9831 

  \\

V1 & 0.6067 & 0.6423 & \textbf{0.8425} & 0.8072 & 0.8498 & \textbf{0.9361} & 0.7491 & 0.7802 & \textbf{0.8221} & 0.9563 & 0.8772 & 0.8998 & \textbf{0.9258} & 0.9785  

  \\

V3 & 0.6551 & 0.6813 & \textbf{0.8167} & 0.8276 & 0.8344 & \textbf{0.9181} & 0.7215 & 0.7412 & \textbf{0.8710} & 0.9678 & 0.9012 & 0.9078 & \textbf{0.9448} & 0.9845 

  \\

V4 & 0.6260 & 0.7083 & \textbf{0.7808} & 0.7732 & 0.8325 & \textbf{0.8913} & 0.5408 & 0.5599 & \textbf{0.6896} & 0.9332 & 0.6922 & 0.7555 & \textbf{0.8484} & 0.9691  

  \\

V5 & 0.6751 & 0.7266 & \textbf{0.8369} & 0.8307 & 0.8722 & \textbf{0.9210} & 0.6190 & 0.6766 & \textbf{0.7250} & 0.9273 & 0.7670 & 0.7820 & \textbf{0.8532} & 0.9661  

  \\

V6 & 0.7276 & 0.7735 & \textbf{0.8580} & 0.8582 & 0.8921 & \textbf{0.9351} & 0.6324 & 0.6358 & \textbf{0.7763} & 0.9334 & 0.8032 & 0.8417 & \textbf{0.8877} & 0.9686  

\\

\hline

Avg  & 0.6934 & 0.7459 & \textbf{0.8484} & 0.8227 & 0.8597 & \textbf{0.9316} & 0.6973 & 0.7505  & \textbf{0.8443} & 0.9501 & 0.8336  & 0.8759 & \textbf{0.9273} & 0.9771  

\\

\hline
\end{tabular}
\label{ptbxl_ptbdb_fulldata}
\end{footnotesize}

\end{center}
\end{table*}

\begin{table*}[t!]
\begin{center}
\caption{Average $R^2$ and $r_x$ between each of the original and reconstructed ECG leads within different patient subgroups in PTB: (a) HC, (b) BB, (c) HY, (d) MI, (e) VA, and (f) ND. Column `a': LSTM model; Column `b': LSTM-UNet model and Column `c': Pix2Pix GAN model in the public setting; and Column `d': LSTM-UNet model in the personalized setting. 
}
\renewcommand{\arraystretch}{1.2}
\renewcommand{\tabcolsep}{3pt}
\begin{tabular}{cc}
{
\begin{footnotesize}
\begin{tabular}{|c|cccc|cccc|}
\hline

\cline{2-9}
& \multicolumn{4}{c|}{$R^2$}
& \multicolumn{4}{c|}{$r_x$} \\
\cline{2-9}

Lead & a & b & c & d & a & b & c & d \\
\hline
III &  0.6479 & 0.6992 & \textbf{0.8913} & 0.9622 & 0.8428 & 0.8459 & \textbf{0.9494} & 0.9817
 \\

avR & 0.9414 & 0.9698 & \textbf{0.9879} & 0.9925 & 0.9711 & 0.9868 & \textbf{0.9961} & 0.9963
  \\

avL & 0.713 & 0.7331 & \textbf{0.8673} & 0.9706 & 0.7441 & 0.8768 & \textbf{0.9447} & 0.9861
  \\

avF & 0.8315 & 0.9245 & \textbf{0.9634} & 0.9835 & 0.9088 & 0.9618 & \textbf{0.9877} & 0.992
  \\

V1 & 0.7021 & 0.7750 & \textbf{0.8545} & 0.9783 & 0.8735 & 0.8898 & \textbf{0.9415} & 0.9894
  \\

V3 & 0.7796 & 0.7969 & \textbf{0.8324} & 0.9764 & 0.9094 & 0.9162 & \textbf{0.9261} & 0.9887
  \\

V4 & 0.4724 & 0.5056 & \textbf{0.5836} & 0.9654 & 0.8099 & 0.8285 & \textbf{0.8601} & 0.9835
  \\

V5 & 0.7111 & 0.7935 & \textbf{0.8790} & 0.9563 & 0.9034 & 0.9101 & \textbf{0.9450} & 0.9796
  \\

V6 & 0.7499 & 0.8407 & \textbf{0.9185} & 0.9762 & 0.9236 & 0.9332 & \textbf{0.9633} & 0.9900
  \\

\hline

Avg & 0.7276 & 0.7820 & \textbf{0.8642} & 0.9734 & 0.8762  & 0.9054 & \textbf{0.9460} & 0.9874 \\

\hline
\end{tabular}
\end{footnotesize}
}
&
{
\begin{footnotesize}
\begin{tabular}{|c|cccc|cccc|}
\hline

\cline{2-9}
& \multicolumn{4}{c|}{$R^2$}
& \multicolumn{4}{c|}{$r_x$} \\
\cline{2-9}

Lead & a & b & c & d & a & b & c & d \\
\hline
III & 0.7029 & 0.7971 & \textbf{0.9002} & 0.9132 & 0.7984 & 0.8502 & \textbf{0.9518} & 0.9613
  \\

avR & 0.8447 & 0.9181 & \textbf{0.9603} & 0.9678 & 0.9207 & 0.9583 & \textbf{0.9832} & 0.9885
  \\

avL & 0.7408 & 0.8414 & \textbf{0.9370} & 0.9599 & 0.8872 & 0.9334 & \textbf{0.9731} & 0.9799
  \\

avF & 0.8328 & 0.9114 & \textbf{0.9684} & 0.9744 & 0.9158 & 0.9593 & \textbf{0.9885} & 0.9944
  \\

V1 & 0.7098 & 0.8060 & \textbf{0.8487} & 0.9533 & 0.8047 & 0.9012 & \textbf{0.9378} & 0.9767
  \\

V3 & 0.8376 & 0.8667 & \textbf{0.9031} & 0.9731 & 0.9330 & 0.9499 & \textbf{0.9617} & 0.9870
  \\

V4 & 0.5380 & 0.6342 & \textbf{0.7012} & 0.9380 & 0.7210 & 0.8058 & \textbf{0.8252} & 0.9700
  \\

V5 & 0.4809 & 0.6862 & \textbf{0.8131} & 0.8996 & 0.7723 & 0.8148 & \textbf{0.8985} & 0.9544
  \\

V6 & 0.5769 & 0.6099 & \textbf{0.7851} & 0.9019 & 0.7508 & 0.7566 & \textbf{0.8692} & 0.9594
  \\

\hline

Avg & 0.6960 & 0.7856 & \textbf{0.8686} & 0.9423 & 0.8337 & 0.8810 & \textbf{0.9321} & 0.9746 \\

\hline
\end{tabular}
\end{footnotesize}
} \\
(a) Healthy controls HC & (b) Bundle branch block BB\\
{
\begin{footnotesize}
\begin{tabular}{|c|cccc|cccc|}
\hline

\cline{2-9}
& \multicolumn{4}{c|}{$R^2$}
& \multicolumn{4}{c|}{$r_x$} \\
\cline{2-9}

Lead & a & b & c & d & a & b & c & d \\
\hline
III & 0.6051 & 0.7482 & \textbf{0.9153} & 0.9611 & 0.8399 & 0.8853 & \textbf{0.9656} & 0.981
  \\

avR & 0.8275 & 0.9346 & \textbf{0.9641} & 0.9801 & 0.9141 & 0.9669 & \textbf{0.9882} & 0.9898
  \\

avL & 0.7459 & 0.7739 & \textbf{0.8860} & 0.9658 & 0.8809 & 0.8984 & \textbf{0.9508} & 0.983
  \\

avF & 0.8028 & 0.9115 & \textbf{0.9132} & 0.9742 & 0.9021 & 0.9564 & \textbf{0.9683} & 0.9874
  \\

V1 & 0.6973 & 0.7157 & \textbf{0.7545} & 0.9712 & 0.8485 & 0.9059 & \textbf{0.9533} & 0.9859
  \\

V3 & 0.6969 & 0.7855 & \textbf{0.8480} & 0.9787 & 0.8240 & 0.9030 & \textbf{0.9344} & 0.9898
  \\

V4 & 0.2854 & 0.3693 & \textbf{0.4980} & 0.9407 & 0.5728 & 0.6339 & \textbf{0.7058} & 0.9711
  \\

V5 & 0.5918 & 0.6377 & \textbf{0.7281} & 0.9241 & 0.6380 & 0.7344 & \textbf{0.8224} & 0.9644
  \\

V6 & 0.4702 & 0.5364 & \textbf{0.7420} & 0.9454 & 0.7381 & 0.8134 & \textbf{0.8731} & 0.9747

  \\

\hline

Avg & 0.6358 & 0.7125 & \textbf{0.8054} & 0.9601 & 0.7953 & 0.8552 & \textbf{0.9069} & 0.9807 \\

\hline
\end{tabular}
\end{footnotesize}
}
&
{
\begin{footnotesize}
\begin{tabular}{|c|cccc|cccc|}
\hline

\cline{2-9}
& \multicolumn{4}{c|}{$R^2$}
& \multicolumn{4}{c|}{$r_x$} \\
\cline{2-9}

Lead & a & b & c & d & a & b & c & d \\
\hline
III & 0.6727 & 0.7354 & \textbf{0.8863} & 0.9161 & 0.8282 & 0.8776 & \textbf{0.9510} & 0.9606
  \\

avR & 0.7976 & 0.9445 & \textbf{0.9705} & 0.9746 & 0.8947 & 0.972 & \textbf{0.9907} & 0.9973
  \\

avL & 0.7517 & 0.8456 & \textbf{0.9324} & 0.9504 & 0.8825 & 0.9255 & \textbf{0.9727} & 0.9786
  \\

avF & 0.7487 & 0.8279 & \textbf{0.9287} & 0.9422 & 0.8769 & 0.9175 & \textbf{0.9754} & 0.9825
  \\

V1 & 0.7381 & 0.7597 & \textbf{0.8182} & 0.946 & 0.8701 & 0.8929 & \textbf{0.9206} & 0.9734
  \\

V3 & 0.7064 & 0.7129 & \textbf{0.8809} & 0.9587 & 0.8093 & 0.9015 & \textbf{0.9498} & 0.9800
  \\

V4 & 0.5300 & 0.6112 & \textbf{0.7241} & 0.9044 & 0.6578 & 0.727 & \textbf{0.8574} & 0.9557
  \\

V5 & 0.4042 & 0.4585 & \textbf{0.7027} & 0.9020 & 0.7606 & 0.7741 & \textbf{0.8427} & 0.9544
  \\

V6 & 0.6099 & 0.6279 & \textbf{0.7587} & 0.9297 & 0.8303 & 0.8340 & \textbf{0.8779} & 0.9583
  \\

\hline

Avg & 0.6621 & 0.7248 & \textbf{0.8447} & 0.9360 & 0.8233  & 0.8691 & \textbf{0.9265} & 0.9712 \\

\hline
\end{tabular}
\end{footnotesize}
}
\\
(c) Hypertrophy, cardiomyopathy and heart failure HY & (d) Myocardial infarction MI \\
{
\begin{footnotesize}
\begin{tabular}{|c|cccc|cccc|}
\hline

\cline{2-9}
& \multicolumn{4}{c|}{$R^2$}
& \multicolumn{4}{c|}{$r_x$} \\
\cline{2-9}

Lead & a & b & c & d & a & b & c & d \\
\hline
III & 0.6218 & 0.7199 & \textbf{0.7589} & 0.9285 & 0.7824 & 0.863 & \textbf{0.8804} & 0.9655
  \\

avR & 0.9638 & 0.9776 & \textbf{0.9877} & 0.9901 & 0.9818 & 0.9896 & \textbf{0.9952} & 0.9958
  \\

avL & 0.677 & 0.7423 & \textbf{0.9262} & 0.9398 & 0.8491 & 0.8765 & \textbf{0.9690} & 0.9707
  \\

avF & 0.6808 & 0.9341 & \textbf{0.9457} & 0.9695 & 0.9136 & 0.9666 & \textbf{0.9771} & 0.9853
  \\

V1 & 0.8898 & 0.9288 & \textbf{0.9417} & 0.9587 & 0.9472 & 0.9656 & \textbf{0.9796} & 0.9797
  \\

V3 & 0.5801 & 0.6591 & \textbf{0.8292} & 0.9753 & 0.7304 & 0.9000 & \textbf{0.9377} & 0.9883
  \\

V4 & 0.5305 & 0.6260 & \textbf{0.7722} & 0.9546 & 0.7037 & 0.7831 & \textbf{0.8816} & 0.9836
  \\

V5 & 0.2726 & 0.3530 & \textbf{0.3809} & 0.9704 & 0.6111 & 0.7275 & \textbf{0.7928} & 0.986
  \\

V6 & 0.8170 & 0.8369 & \textbf{0.9052} & 0.9788 & 0.9089 & 0.9289 & \textbf{0.9586} & 0.9857
  \\

\hline

Avg & 0.6703 & 0.7530 & \textbf{0.8275} & 0.9628 & 0.8253  & 0.8889 & \textbf{0.9302} & 0.9822 \\

\hline
\end{tabular}
\label{incart_results}
\end{footnotesize}
}
&
{
\begin{footnotesize}
\begin{tabular}{|c|cccc|cccc|}
\hline

\cline{2-9}
& \multicolumn{4}{c|}{$R^2$}
& \multicolumn{4}{c|}{$r_x$} \\
\cline{2-9}

Lead & a & b & c & d & a & b & c & d \\
\hline
III & 0.7167 & 0.8075 & \textbf{0.9019} & 0.9033 & 0.8616 & 0.9191 & \textbf{0.9584} & 0.9561
  \\

avR & 0.8330 & 0.9406 & \textbf{0.9745} & 0.9773 & 0.9134 & 0.9705 & \textbf{0.9907} & 0.9992
  \\

avL & 0.6709 & 0.8760 & \textbf{0.8810} & 0.9526 & 0.8348 & 0.9319 & \textbf{0.9391} & 0.9770
  \\

avF & 0.7871 & 0.9146 & \textbf{0.9482} & 0.9537 & 0.8960 & 0.9617 & \textbf{0.9846} & 0.9879
  \\

V1 & 0.6421 & 0.7574 & \textbf{0.7816} & 0.9306 & 0.8466 & 0.8902 & \textbf{0.8910} & 0.9660
  \\

V3 & 0.7117 & 0.8033 & \textbf{0.8588} & 0.9451 & 0.8857 & 0.9201 & \textbf{0.9281} & 0.9735
  \\

V4 & 0.4898 & 0.5988 & \textbf{0.6884} & 0.8962 & 0.6889 & 0.8050 & \textbf{0.8296} & 0.9510
  \\

V5 & 0.3794 & 0.4560 & \textbf{0.5409} & 0.9114 & 0.5579 & 0.6052 & \textbf{0.6865} & 0.9578
  \\

V6 & 0.3296 & 0.3956 & \textbf{0.4944} & 0.8785 & 0.7127 & 0.7335 & \textbf{0.7632} & 0.9435
  \\

\hline

Avg & 0.6178 & 0.7277 & \textbf{0.7855} & 0.9276 & 0.7997  & 0.8596 & \textbf{0.8857} & 0.9680 \\

\hline
\end{tabular}
\label{incart_results}
\end{footnotesize}
}
\\
(e) Valvular myocarditis and miscellaneous VA & (f) Without diagnostic data ND \\
\end{tabular}
\label{ptbdb_grouped}

\end{center}
\end{table*}

\subsection{Proposed workflow and performance measures}



The Pix2Pix GAN model was implemented using the PyTorch library. The UNet generator consists of 7 downsampling and upsampling blocks which make up the encoder and decoder respectively. Each block consists of a 1D convolution/transpose convolution layer, a batch normalisation layer and an activation function. For downsampling, we use the Leaky Rectified Linear Unit (LeakyReLU) activation and for upsampling regular ReLU is used. For the last upsampling block, we use $\tanh$ activation to ensure the output lies in the range [-1,1]. The input to the model consists of 3 channels of 1D data of length $L$, i.e., a $3\times L$ matrix, which is reconstructed to form a $9\times L$ matrix, where $L$ is the length of the signal cropped to the nearest multiple of $128$. The output shares the same dimensions. During training, we used training samples of size $3 \times 4992$. In case of longer sequences, multiple training sequences were obtained via a sliding window technique. This ensured a fixed signal length for use with PyTorch tensors and dataloaders. For evaluation, we supplied the original length of the signal to the model without windowing. The model was trained for 14 epochs, with early stopping based on the highest $R^2$ score on the test dataset and a patience value of 3 epochs.


As measures of ECG signal reconstruction performance/accuracy, we made use of coefficient of determination ($R^2$) and Pearson's correlation coefficient ($r_x$). Specifically, considering patient record $x_i$, $i=1,2,...,n$, reconstructed to $y_i$, the aforesaid performance indices are computed by $R^2 = 1-\frac{\Sigma_{i=1}^n {(x_i-\bar{y})}^2}{\Sigma_{i=1}^n {(y_i-\bar{y})}^2}$ and $r_x = \frac{\Sigma_{i=1}^n {(x_i-\bar{x})} {(y_i-\bar{y})}}{\sqrt{\Sigma_{i=1}^n {(x_i-\bar{x})}^2} \sqrt{\Sigma_{i=1}^n {(y_i-\bar{y})}^2}}$. A perfect reconstruction would achieve $R^2 = 1$ (within the range $[0,1]$) and $r_x = 1$ (within the range $[-1,1]$).


\section{Results}

For visual appreciation of the reader, we begin by presenting in Figure \ref{HC_UH} the waveforms of the nine reconstructed leads (namely, III, aVR, aVL, aVF, V1, V3, V4, V5, and V6) obtained using the 1D Pix2Pix GAN model in the public setting (in red, whereas the original is in blue) based on three given leads (I, II and V2) as input for two representative records (Record No: patient195/s0337 in left panel, Record No: patient280/s0535 in right panel, MI sub-group patients from PTB). As a state-of-the-art reference, the same reconstructions performed by LSTM-UNet in the personalized setting were depicted in green. Except for leads V4, V5 and V6, the public setting reconstructions appear reasonably close to reference personalized ones.


Next, we statistically analyzed the test performance of the proposed method, and furnished in Table \ref{ptbxl_ptbdb_fulldata}. On the PTBXL database, in the public setting, the proposed GAN model achieves an overall $R^2$ (resp. $r_x$) measure of 84.84\% (resp. 93.16\%), which improves upon existing LSTM and LSTM-UNet performances of 69.34\% (resp. 82.27\%) and 74.59\% (resp. 85.97\%), respectively. Inspecting lead-wise statistics closely, leads such as avR ($R^2=$97.58\%, $r_x=$99.17\%) appeared to account for fairly accurate reconstruction, whereas leads such as V4 ($R^2=$78.08\%, $r_x=$89.13\%) appeared less satisfactory. 
On the PTB database, again in the public setting, the aforementioned algorithms exhibited performances that were broadly similar, numerically. However, we now had an additional reference of personalized performance available from another source \cite{lstm-unet}. The $R^2$ (resp. $r_x$) measure of 84.43\% (resp. 92.73\%) in the public setting compares with 95.01\% (resp. 97.71\%) in the personalized case, indicating a still significant gap. However, for certain leads such as avR, the said gap is less significant (97.29\% versus 98.20\% in terms of $R^2$, and 99.12\% versus 99.77\% in terms of $r_x$), whereas for certain other leads, such as V4, such gap is more prominent (68.96\% versus 93.32\% in terms of $R^2$, and 84.84\% versus 96.91\% in terms of $r_x$).


Continuing with the PTB database, we next performed similar statistical analysis within patient subgroups, namely, healthy HC, and diseased BB, HY, MI, VA, and ND, and furnished in Table \ref{ptbdb_grouped}. In the public setting, the proposed GAN model consistently outperformed competing models in terms of both $R^2$ and $r_x$. Further, the performance gap between the public and the personalized settings appeared to be less in the healthy control (HC), bundle branch block (BB), and myocardial infarction (MI) subgroups, and somewhat more in hypertrophy, cardiomyopathy and heart failure (HY) and valvular myocarditis and miscellaneous (VA) subgroups. The gap was the most significant in the default subgroup ND (without diagnostic data), but as clinical significance cannot be assigned there, it was presently ignored. In most subgroups, a few leads, such as V4, V5 and V6, proved intransigent towards reconstruction in the public setting. The above identifies the contexts (given by specific subgroups and specific leads) where the current AI model needs significant improvement before becoming practically useful. 

\section{Discussion}

In this paper, we considered 3-lead to 12-lead ECG reconstruction in the public setting, which, as mentioned, has hitherto received scant research attention. There, a 1D Pix2Pix GAN model was found to provide performance improvement on multiple test datasets over LSTM and LSTM-UNet models that proved efficient in the personalized setting. While this marks an encouraging progress, the proposed model should be considered for practical deployment only upon significant further improvement, especially in some underperforming ones of the chest leads V1--V6 (which are crucial in the diagnosis of important cardiac conditions including MI). In general, ECG being a primary tool for diagnosing critical cardiac conditions, even seemingly minor inaccuracies in lead reconstruction could lead to misdiagnosis, potentially endangering human lives. Indeed, it is imperative to ensure algorithmic robustness and reliability before proceeding to clinical trials. At the same time, continued research in this direction should be useful because, if/when proven satisfactory, the GAN model under consideration will have practical relevance in smartphone applications due to its compact architecture (with only about 30 million parameters). The resulting setup, with real-time on-device processing and without needing external computational resources, could be attractive in the target remote healthcare scenarios.

\section{Future work}

A related problem of ECG signal synthesis arises from the growing need for statistically authentic synthetic data for testing medical devices and algorithms. Variations of GAN models, such as time-series GAN and LSTM-GAN \cite{gantimeseries, lstmgan}, have proven useful in the above  and related problems. Whether such GAN variants are efficient for our lead reconstruction problem should be a worthwhile exploration as well. Additionally, the cost function can be made dependent on known lead information for better contextualization and potentially improved efficacy. We also envisage a further study of whether the use of side information such as a (cheaply acquired) photoplethysmograph (PPG) channel alongside the 3-lead ECG data enhances robustness. Of course, a novel approach could also consist in algorithmically generating health alerts based on three-lead ECG data (refer to the dashed path in Figure \ref{general_monitoring}(b)), and undertaking the present lead reconstruction task only upon alert, thus saving resources.


\begin{thebibliography}{00}

\bibitem{12lead} A.H. Ribeiro \textit{et al}., ``Automatic diagnosis of the 12-lead ECG using a deep neural network,'' \textit{Nat. Commun.}, vol. 11, no. 1, p. 1760, 2020.


\bibitem{R3lead} J.A. Scherer, J. M. Jenkins, and J. M. Nicklas, ``Synthesis of the 12-lead electrocardiogram from a 3-lead subset using patient-specific transformation vectors—An algorithmic approach to computerized signal synthesis,'' \textit{J. Electrocardiol.}, vol. 22, Suppl, pp. 128-128, 1989.

\bibitem{dower} G.E. Dower., ``A lead synthesizer for the Frank system to simulate the standard 12-lead electrocardiogram.'' \textit{J. Electrocardiol.}, vol. 1, no. 1, pp. 101--116, 1968.

\bibitem{affine} D. Dawson, H. Yang, M. Malshe, ST Bukkapatnam, B. Benjamin and R. Komanduri., ``Linear affine transformations between 3-lead (Frank XYZ leads) vectorcardiogram and 12-lead electrocardiogram signals,'' \textit{J. Electrocardiol.}, vol. 42, no. 6, pp. 622–630, 2009.


\bibitem{amit} S. Maheshwari, A. Acharyya, P. Rajalakshmi, P.E. Puddu, M. Schiariti., ``Accurate and reliable 3-lead to 12-lead ECG reconstruction methodology for remote health monitoring applications,'' \textit{IRBM}, vol. 35, no. 6, pp. 341--350, 2014.

\bibitem{lstm} J. Sohn, S. Yang, J. Lee, Y. Ku, H.C. Kim., ``Reconstruction of 12-lead electrocardiogram from a three-lead patch-type device using a LSTM network,'' \textit{Sensors}, vol. 20, no. 11, pp. 3278, 2020.

\bibitem{lstm-unet} L.R. Rahul, A. Shaiju and S. Jana., ``3-Lead to 12-Lead ECG Reconstruction: A Novel AI-based Spatio-Temporal Method,'' \textit{20th INDICON}, pp. 957--962, 2023.

\bibitem{physionet} G.B. Moody, R.G. Mark, and A.L. Goldberger, ``PhysioNet: a web-based resource for the study of physiologic signals,'' \textit{IEEE Eng. Med. Biol. Mag.}, vol. 20, no. 3, pp. 70--75, 2001.

\bibitem{sandeep} B.S. Chandra, C.S. Sastry, S. Jana., ``Robust heartbeat detection from multimodal data via CNN-based generalizable information fusion,'' \textit{IEEE Trans. Biomed. Eng.}, vol. 66, no. 3, pp. 710--717, 2020.

\bibitem{ganconditional} P. Isola, J.Y. Zhu, T. Zhou and A.A. Efros., ``Image-to-image translation with conditional adversarial networks,'' \textit{CVPR}, pp. 1125--1134, 2017. 


\bibitem{gantimeseries}  R. Banerjee,  O. Mazumder, A. Mukherjee,  S. Sinhahajari and A. Sinha., ``Reconstruction of Body Surface Potential From 12-Lead ECG: A Conditional GAN Based Approach,'' \textit{31st EUSIPCO}, pp. 1180-1184, 2023.

\bibitem{lstmgan} Y.H. Zhang and S. Babaeizadeh., ``Synthesis of standard 12‑lead electrocardiograms using two-dimensional generative adversarial networks,'' \textit{J. Electrocardiol.}, vol. 69, pp. 6--14, 2021.

\end{thebibliography}
\end{document}